\documentclass[a4paper,twoside,10pt]{article}
\usepackage{graphicx,publaob}

\def\papertype{Contributed paper}
\begin{document}

\title{SCALING OF THE DIFFUSION COEFFICIENT ON THE NORMAL FORM 
REMAINDER IN DOUBLY-RESONANT DOMAINS}

\authors{C. EFTHYMIOPOULOS$^1$}

\address{$^1$Research Center for Astronomy, Academy of Athens, 
Soranou Efessiou 4, 115 27 Athens, Greece}
\Email{cefthim}{academyofathens}{gr}

\markboth{DIFFUSION IN DOUBLY-RESONANT DOMAINS}{C. EFTHYMIOPOULOS}

\abstract{An outline of theoretical estimates is given regarding  
the dependence of the value of the diffusion coefficient $D$ 
on the size $R$ of the remainder of the normal form in doubly 
or simply resonant domains of the action space of 3dof
Hamiltonian systems.}

\section{INTRODUCTION}

In the works of  Froeschl\'{e} et al. (2000, 2005, 2006), Lega et al. (2003), 
and Guzzo et al. (2005) precise numerical estimates were 
given of the critical threshold (in terms of a small parameter $\epsilon$) 
at which one has the onset of the so-called `Nekhoroshev regime' 
(Nekhoroshev 1977, Benettin et al. 1985, Lochak 1992, P\"{o}shel 1993) 
in conservative systems of three degrees of freedom. In the same works 
the local speed of Arnold diffusion, as well as the exponents appearing 
in the associated laws of the Nekhoroshev theory, were estimated.
Other studies by the same group are Guzzo et al. (2006, effects 
of non-convexity) and Todorovic et al.(2008, diffusion in `a priori 
unstable' systems). Earlier studies are: Kaneko and Konishi (1989), 
Wood et al. (1990), Dumas and Laskar (1993), Laskar (1993), Skokos 
et al. (1997), Giordano and Cincotta (2004).

In a recent study (Efthymiopoulos 2008), we used a computer program to 
carry on Hamiltonian normalization up to a sufficiently high order, and 
found that in a case of simple resonance the size $R$ of the optimal 
remainder of the normalized Hamiltonian (which turned to be exponentially small 
in $1/\epsilon^a$, $a\simeq 0.25$) scales with the diffusion coefficient 
of numerical experiments as $D\propto R^3$. This scaling law was presented
as an empirical fact, no theory behind being suggested. Furthermore, no 
investigation was made of what happens in cases of double resonance. 
The above two questions are briefly addressed in the sequel. In particular, 
theoretical estimates of the scaling law $D(R)$ are outlined, first in 
the case of double and then of simple  resonances. Details are deferred 
to a paper under preparation.

\section{OUTLINE OF ESTIMATES ON THE  LAW $D(R)$}

We shall refer to Hamiltonian systems of three degrees of freedom 
\begin{equation}\label{hamgen}
H(I,\phi)=H_0(I)+\epsilon H_1(I,\phi)
\end{equation}
where $I\equiv(I_1,I_2,I_3)$, $\phi\equiv(\phi_1,\phi_2,\phi_2)$, 
$H_0$ satisfies appropriate non-degeneracy and convexity conditions, 
and $H$ is analytic in a complexified domain of the actions and 
the angles. The frequencies are $\omega(I)=\nabla_I H_0(I)$. 
Owing to the analyticity condition, the Fourier series 
$H_1=\sum_k h_k(I)\exp(ik\cdot\phi)$
yields coefficients $h_k$ decaying exponentially with the modulus 
of the vector $k$, namely the bound $|h_k(I)|< A\exp(-|k|\sigma)$
holds with $|k|=|k_1|+|k_2|+|k_3|$ and $A,\sigma$ positive constants. 
Consider an open domain ${\cal W}_{I_*,D}$ in the action space, 
of size $D$, centered around some central value $I_*$. Expanding 
$H_0$ around $I_*$ leads to 
\begin{equation}\label{h0exp}
H_0=H_{0*}+\omega_*\cdot J + {1\over 2}M_{ij*}J_iJ_j +\ldots
\end{equation}
where $J=I-I_*$, $\omega_*=\nabla_I H_0(i_*)$, $M_{ij*}$ are the 
entries of the Hessian matrix of $H_0$ at $I_*$ (which satisfy 
restrictions imposed by the convexity condition).
Assuming some truncation order $K_c$ in Fourier space, we consider 
the case in which there are two linearly independent integer vectors 
$k^{(1)}$, $k^{(2)}$ both satisfying the conditions $|k^{(j)}|<K_c$, 
$k^{(j)}\cdot\omega_*=0$, $j=1,2$. A normalization of the Hamiltonian 
(\ref{hamgen}) valid in ${\cal W}_{D,I_*}$ results in that all the 
resonant trigonometric terms, of the form
$\exp(i((q_1k^{(1)}+q_2k^{(2)})\cdot\phi)$, with $q_1,q_2$
integers, $|q_1k^{(1)}+q_2k^{(2)})|<K_c$, survive in the
normal form. Choosing a vector m satisfying $m\cdot k^{(j)}=0, 
j=1,2$, and setting $J_i=k_i^{(1)}I_R^{(1)}+k_i^{(2)}I_R^{(2)}+m_iI_F$,
$\phi_R^{(j)}=k^{(j)}\cdot\phi$, $\phi_F=m\cdot\phi$, $i=1,2,3$,
$j=1,2$, the normalized Hamiltonian takes the form (in new canonical variables)
\begin{eqnarray}\label{hamnf}
H_{norm} =Z_R\left(I_R^{(1)},I_R^{(2)},I_F,\phi_R^{(1)},\phi_R^{(2)}\right)+
R\left(I_R^{(1)},I_R^{(2)},I_F,\phi_R^{(1)},\phi_R^{(2)},\phi_F\right)~.
\end{eqnarray}
Two facts are relevant about (\ref{hamnf}): i) the angle $\phi_F$
is ignorable in the normal form $Z_R$. Therefore $I_F$ is an
integral of the Hamiltonian flow under $Z_R$ alone. It follows
that, for different label values of $I_F$, $Z_R$ alone defines
the dynamics of a system of two degrees of freedom. ii) The
(optimal) remainder is exponentially small $R=O(\exp(-1/\epsilon))$,
thus it only slightly alters the dynamics due to the normal form.

Now, the convexity condition ensures that there is a linear canonical 
transformation $(J_R^{(1)},J_R^{(2)},\phi_R^{(1)},\phi_R^{(2)})$
$\rightarrow$ $(J'_1,J'_2,\phi'_1,\phi'_2)$
such that in the new variables the Hamiltonian reads 
$Z_R ={1\over 2}\left[{\cal A}_1(J_1'-c_{F,1})^2
+{\cal A}_2(J_2'-c_{F_2})^2\right]+\ldots
+\epsilon\sum_{n}C_n(J_1',J_2';\epsilon,I_F)$ $\times\exp(in\cdot\phi')$, 
the constants ${\cal A}_1$, ${\cal A}_2$ being positive and depending on 
$M_{ij*}$, and $c_{F,1},c_{F,2}$ depending on $M_{ij*}$ as well as on the 
value of the integral $I_F$. Apart from trivial modifications, 
the dynamics of $Z_R$ is then the same as under the simplified model
\begin{equation}\label{nfsmp}
Z_R ={1\over 2}\left[(J_1'-c_{F,1})^2
+(J_2'-c_{F_2})^2\right]+
\epsilon\sum_{n}C_n(J_1',J_2';\epsilon,I_F)\exp(in\cdot\phi')
\end{equation}
which we now consider in some detail. 

For any fixed value of the angles $(\phi_1',\phi_2')$, the constant 
energy condition $Z_R=E$ implies that the actions $J_1',J_2'$ lie on 
a circle of radius $\sqrt{2E}$ centered at $(J_1',J_2')=(c_{F,1},c_{F_2})$. 
This is shown schematically in Figure 1. As the angles $\phi'$ change 
value, the trigonometric terms in (\ref{nfsmp}) can only induce 
a $O(\epsilon)$ change in the radius of the circle, which
we shall temporarily ignore. On the other hand, 
the lines $m_1'J_1'+m_2'J_2'=0,~~~m'\cdot n = 0$, 
where $n$ is the integer vector of any trigonometric term 
appearing in (\ref{nfsmp}), define the 1D resonant manifolds of 
the Hamiltonian (\ref{nfsmp}) in the reduced 2D action space. 
By construction, the Hamiltonian (\ref{nfsmp}) contains only a 
finite number of harmonics with coefficients $C_n$ scaling as 
$|C_n|\sim \epsilon\exp(-\sigma |n|)$. It follows that there is 
a finite set of resonant lines in Fig.1, all passing from the 
center of the circles of constant energy. Three such lines are shown 
schematically (black solid). The pairs of dashed lines defining 
the boundaries of the zones around the resonant lines correspond 
to the limits of the separatrix width of each resonance, which is 
of the order 
\begin{equation}\label{sepw}
\Delta J_n\sim \left(\epsilon\exp(-\sigma |n|)\right)^{1/2}~~.
\end{equation}
The smaller the value of $|n|$ the larger the width of the 
resonance. Thus, in Fig.1 we have $|n_1|<|n_2|<|n_3|$. 

The resonances are well separated if the energy $E$ is {\it large} 
enough (big circle in Fig.1, at $E=E_1$). Since the entire set of 
resonances cover an $O(\epsilon^{1/2})$ length of an arc on a circle 
of constant energies, significant resonance overlap occurs if 
$E=O(\epsilon)$ or smaller (small circle for $E=E_2$ in Fig.1). 

So far we have neglected the effect of the remainder. This is 
manifested by causing a slow drift of the normal form energy value 
$E_{NF}$ of the orbits run under the full Hamiltonian, compensating 
the drift in the remainder value so that the total energy $E=E_{NR}+R$ 
be a constant. In Fig.1 this drift is shown by curly curves. An orbit
found at a given time near the separatrix of, say, resonance 1, 
with $E_{NF}=E_1$, has a certain probability to slowly drift in the action 
space outwards ($E_{NF}$ increases) or inwards ($E_{NF}$ decreases). 
Assume the latter case. Then, after a long time the orbit will be 
found touching the inner circle $E=E_2$, where all resonances 
overlap. If, still later, the drift is reversed (again with a 
certain probability), $E_{NF}$ will increase and the orbit will 
return to the circle $E_{NF}=E_1$, but this time lying, in general, 
on the separatrix layer of a {\it different} resonance, say resonance 
2.  

This qualitative picture of the diffusion in resonance junctions should, 
of course, be substantiated by calculating the {\it invariant manifolds} 
of the 2D tori lying at the borders of the resonances of the 
full Hamiltonian system. If there are heteroclinic intersections 
between these manifolds, an orbit started on one manifold is 
guaranteed to undergo the type of drift motion described above. 
However, verifying this fact numerically poses a formidable 
task, equivalent to probing the very mechanism of Arnold 
diffusion. This restriction notwithstanding, we proceed in 
a plausible quantitative estimate of the speed of the diffusion 
on the basis of numerical experiments having observed, instead, 
the drift in the action space directly (as e.g. in Lega et al. 
2003). The indication from such experiments is that the drift 
in the action space can be modeled, at least locally, like a 
normal diffusion process. Assuming this true, one may ask how 
long it will take for, say, the curly itinerary of Fig.1 to be 
realized. Attributing the drift to the remainder, the per step 
change of the $E_{NF}$ value of an orbit crossing the junction 
is $dE\sim R$. By Fick's law, one then has after $N$ steps a 
total change $\Delta E_{NF}\sim N^{1/2} dE=N^{1/2} R$. 
By (\ref{sepw}) the radius of the outermost circle touching the 
junction's limits is $\Delta J=O(\epsilon^{1/2})$, implying that 
the total drift in the value of $E_{NF}$ is $\Delta E_{NF}=O(\epsilon)$. 
Putting these relations together one then finds  for the diffusion 
coefficient $D\sim\Delta J^2/N$ the relation $D\sim \epsilon^{-1} R^2$.
Note that, since $R=O(\exp(-1/\epsilon))$ one has, by 
de l'Hopital's rule that $D\sim R^2\rightarrow 0$ in the 
limit $\epsilon\rightarrow 0$. 
\begin{figure}
\centering
\includegraphics[width=8cm,height=5.5cm]{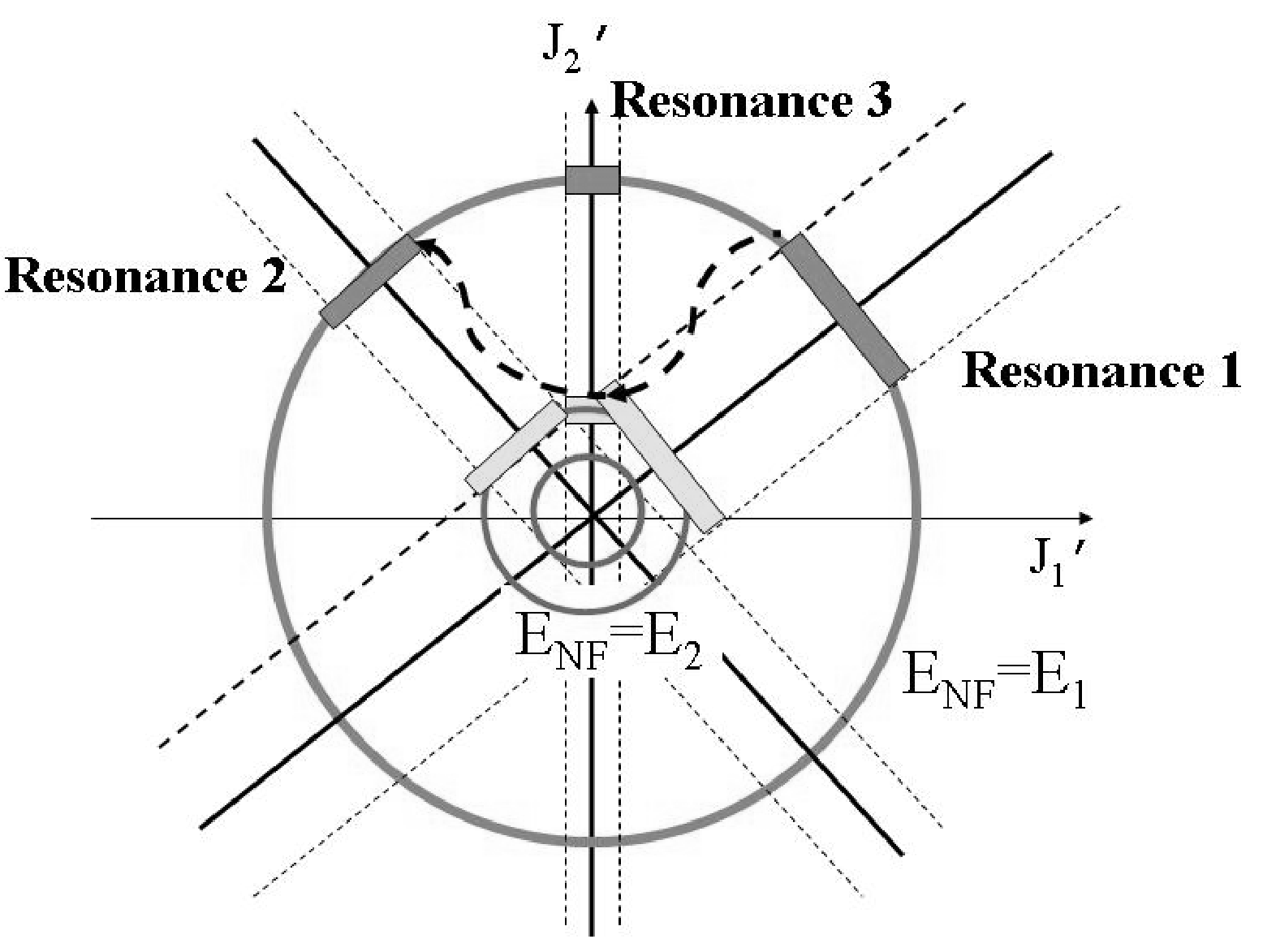}
\caption{The circles of constant normal form energy $E_{NF}$ and 
the drift of the orbits in resonance junctions (see text)}
\end{figure}

A similar calculation can be made in the case of simple resonances, 
if we take into account that the simply-resonant domains are, in fact, 
composed also by sub-domains ${\cal W}_{D,I_*}$ crossed by multiple 
resonances, the difference from doubly resonant domains being essentially 
that only one of these resonances satisfies $|k^{(1)}|<K_c$, while all 
the others satisfy $|k^{(2)}|\geq K_c$. Since all apart one resonances  
are contained in the remainder function, it follows that the total 
travel to cross a resonant junction in the action space is now of 
order $R^{1/2}$ (instead of $\epsilon^{1/2}$ in double resonances). 
Setting $\Delta J^2\sim R$ we then readily find $D\sim \epsilon^{-2} R^3$.

In conclusion, we predict that in systems like (\ref{hamgen}) 
{\it the local value of the diffusion 
coefficient $D$ scales with the size of the optimal remainder function 
$R$ of the local normal form as a power law $D\sim R^{b}$}, the estimate
$b\approx 2$ holding close to doubly-resonant domains and 
$b\approx 3$ close to simply-resonant domains. It would be of 
interest to check these theoretical estimates against detailed numerical 
experiments. 

\references

Arnold, V.I., 1964: \journal{Sov. Math. Dokl.} \vol{6}, 581.

Benettin, G., Galgani, L., and Giorgilli, A.: 1985, \journal{Cel. Mech.}
\vol{37}, 1.

Dumas, H.S., and Laskar, J.: 1993, \journal{Phys. Rev. Lett.} \vol{70}, 2975.
Froeschl\'{e}, C., Guzzo, M., and Lega, E.: 2000, \journal{Science} \vol{289}
(5487), 2108.

Efthymiopoulos, C: 2008, \journal{Cel. Mech. Dyn. Astron.}, \vol{102}, 49.

Froeschl\'{e}, C., Guzzo, M., and Lega, E.: 2000, \journal{Science} \vol{289}
(5487), 2108.

Froeschl\'{e}, C., Guzzo, M., and Lega, E.: 2005, \journal{Cel. Mech. Dyn. Astron.}
\vol{92}, 243.

Froeschl\'{e}, C., Lega, E., and Guzzo, M.: 2006, \journal{Cel. Mech. Dyn. Astron.}
\vol{95}, 141.

Giordano, C.M., and Cincotta, P.M.: 2004, \journal{Astron. Astrophys.} \vol{423}, 745.

Guzzo, M., Lega, E., and Froeschl\'{e}, C.: 2005, \journal{Dis. Con. Dyn. Sys. B}
\vol{5}, 687.

Guzzo, M., Lega, E., and Froeschl\'{e}, C.: 2006, \journal{Nonlinearity}, 
\vol{19}, 1049.

Kaneko, K., and Konishi, T.: 1989, \journal{Phys. Rev. A} \vol{40}, 6130.

Laskar, J.: 1993, \journal{Physica} \vol{D67}, 257.

Lega, E., Guzzo, M., and Froeschl\'{e}, C.: 2003, \journal{Physica D} \vol{182}, 179.

Lochak, P.: 1992, \journal{Russ. Math. Surv.} \vol{47}, 57.

Nekhoroshev, N.N.: 1977, \journal{Russ. Math. Surv.} \vol{32}(6), 1.

P\"{o}shel, J.: 1993, \journal{Math. Z.} \vol{213}, 187.

Skokos, C., Contopoulos, G., and Polymilis, C.: 1997, \journal{Cel. Mech. Dyn. Astr.}
\vol{65}, 223.

Todorovic, N., Lega, E., and Froeschl\'{e}, C.: 2008, \journal{Cel. Mech. Dyn. Astr.},
\vol{19}, 1049.

Wood, B.P., Lichtenberg, A.J., and Lieberman, M.A.: 1990,\journal{Phys.
Rev. A} \vol{42}, 5885.

\endreferences

\end{document}